\documentclass[11pt,preprint]{aastex}

\begin{document}

\title{On the observational properties of He-burning stars:
some clues on the tilt of the HB in metal rich clusters}

\author{G. Raimondo\altaffilmark{1,2}, V. Castellani\altaffilmark{3,4}, 
S. Cassisi\altaffilmark{1,5}, E. Brocato\altaffilmark{1,6} and 
G. Piotto\altaffilmark{7}}

\altaffiltext{1}{Osservatorio Astronomico di Collurania, Via M. Maggini, 
I-64100 Teramo, Italy; brocato, cassisi, raimondo@astrte.te.astro.it}

\altaffiltext{2}{Astronomia-Dipartimento di  Fisica, Universit\`a La Sapienza, 
P.le A. Moro 2, I-00185 Roma, Italy}

\altaffiltext{3}{Dipartimento di Fisica, Universit\`a di Pisa, P.za 
Torricelli 2, I-56100 Pisa, Italy; vittorio@astr18pi.difi.unipi.it}

\altaffiltext{4}{Istituto Nazionale di Fisica Nucleare, Sezione di Pisa, 
I-56100 Pisa, Italy}

\altaffiltext{5}{Max Planck Institut f\"{u}r Astrophysik, K. 
Schwarzschild-Strasse 1, 85740 Garching, Munchen, Germany}

\altaffiltext{6}{Istituto Nazionale di Fisica Nucleare, LNGS, 
I-67100 L'Aquila, Italy}

\altaffiltext{7}{Dipartimento di Astronomia, Universit\`a di Padova, 
Vicolo dell'Osservatorio 5, I-35122 Padova, Italy; piotto@pd.astro.it}

\begin{abstract}
We investigate the predicted Color-Magnitude distribution of
metal-rich Horizontal Branch (HB) stars, discussing selected
theoretical models computed under various assumptions about the star
metallicity and the efficiency of super-adiabatic convection.  We find
that canonical Zero Age Horizontal Branches with metallicity larger or
of the order of Z=0.002 should be all affected by a tilt, by an amount
which increases when the metallicity is increased and/or the mixing
length is decreased, reaching a tilt of $\Delta V \sim$0.2 mag in the
case of solar metallicity when a mixing length value $\alpha$=1.6 is
assumed ($\Delta V$ is the magnitude difference between the top of the
blue HB and the fainter magnitude reached by the red HB).
Uncertainties in the luminosity of the red HB due to uncertainty in
the mixing length value are discussed.  We finally discuss the much
larger tilt observed in the clusters NGC~6441 and NGC 6388, reporting
additional evidence against suggested non-canonical evolutionary
scenarios.  Numerical experiments show that differential reddening
could produce such sloped HBs.  Further, HST-PC imaging of NGC~6441
gives clear indications about the occurrence of differential reddening
across the cluster. However, the same imaging shows that the observed
slope of the red HB {\em is not} an artifact of differential
reddening.  We finally show that sloping red HBs in metal rich
clusters are a common occurrence not necessarily correlated with the
appearance of extended blue HB.
\end{abstract}


\section{Introduction} 

In a recent paper Brocato et al. (1999, paper I) we have presented the
case of the intermediate metallicity globular cluster NGC 6362, where
Horizontal Branch (HB) stars reach a maximum V luminosity at
(B$-$V)$\sim$0.2 mag, with an overall tilt of the order of $\Delta
V$$\sim$0.1 mag, thus of the same order of magnitude of the tilted HB
already observed in NGC 1851 (Walker 1998), possibly in NGC 6229
(Borissova et al. 1999) and in NGC 6712 (Ortolani et al. 2000).  In
the same paper it has been shown that canonical predictions concerning
low-mass, He-burning stars with Z= 0.002 do foresee such an
occurrence, nicely accounting for the observed NGC 6362 HB stars
distribution. Thus we regard such a tilt as an evolutionary feature
which should characterize all the well populated intermediate
metallicity clusters when observed with sufficient photometric
accuracy.

The evidence for such a "canonical" tilt, as due to the behavior of
the bolometric correction, appears of some relevance in connection
with the debated problem of the much larger tilts ($\Delta V$$\sim$0.5
mag) observed in some metal rich globulars of the inner Milky Way.
The first evidence for similar unexpectedly large tilts was brought to
the light from HST observations of the globulars NGC~6388 and NGC~6441
(Piotto et al. 1997, Rich et al. 1997). Sweigart and Catelan (1998)
drove the attention on such a feature, regarding it as an
observational evidence requiring non-canonical evolutionary scenarios.
In the present paper we investigate evolutionary predictions
concerning the Color-Magnitude (CM) distribution of metal rich HB
stars. In the next section we will present new sets of canonical HB
models, whereas section 3 will deal with their CM diagram and
theoretical uncertainties, due to the efficiency of the convection in
superadiabatic layers, which affect the predicted luminosity of stars
at the red HB end.  In section 4 we will provide arguments and
constraints to understand the origin of the large tilt of the high
metallicity clusters NGC~6441 and NGC~6388.  Final remarks will close
the paper.


\section {Metal-rich Zero Age Horizontal Branch models}

The recent literature has already devoted considerable attention to
metal rich HB structures. Starting from the pioneering paper by Horch,
Demarque \& Pinsonneault (1992) several works have been presented
covering this issue with rather detailed evaluations (Dorman, Rood \&
O'Connel 1993, Fagotto et al. 1994, Yi et al. 1997, Bono et al. 1997,
Girardi et al. 2000). However, these studies have been mainly devoted
to investigating the evolution of HB structures in connection with
either the UV excess in elliptical galaxies or the pulsational
properties of metal rich RR Lyrae variables and, to our knowledge, up
to recent times no attention was devoted to the predicted distribution
of metal rich HB stars in the CM diagram.  To investigate such an
issue, we decided to extend the evolutionary computations given in
Paper I for Z=0.002 to higher metallicities.  Similar computations
have been recently presented by VandenBerg et al. (2000) but for a
more restricted range of metallicity, not addressing the problem of
tilt.  However, this will allow an useful comparison between the
results of the two theoretical scenarios for a common value of Z.


In this paper, Zero Age Horizontal Branch (ZAHB) models have been
produced by following selected stellar models all along their
H-burning phase till the ignition of the He flash, deriving in this
way the mass of the He core of new born He burning stars with an age
of the order of 10 Gyr, in agreement with several recent estimates of
the age of metal rich GCs (Salaris \& Weiss 1997, Gratton et al. 1997,
Chaboyer et al. 1998, Cassisi et al. 1999).  For a detailed discussion
on the physical inputs adopted in the computations we refer to Cassisi
et al. (1998).  Table~1 gives selected parameters for the above quoted
models.  Left to right one finds: original He abundance, metallicity
and mass of the model, mass of the He core at the He ignition, surface
He abundance after the first dredge up and, finally, star age and
luminosity at the RGB tip.

As shown in Table 1, to extend the investigation in Paper I at higher
metallicities, we selected two "suitable" values of Z, choosing
Z=0.006 as representative of "metal rich" clusters like 47~Tuc and
Z=0.02 as a safe upper limit for globular cluster metallicities.  As
in Paper I, the present computations have been performed by adopting a
mixing length parameter $\alpha$=1.6, implemented with the case
$\alpha$=1.0 to explore the effect of mixing length assumptions.
Figure~1 gives the predicted distribution of these models in the
log(L/L$_{\odot})$, logT$_{e}$ HR diagram. As labelled in the figure,
the amount of original He (Y$_{MS}$) has been increased with Z
according to $\delta$Y/$\delta$Z $\sim$2.5, as suggested by current
evaluation on that matter (see, e.g., Peimbert 1995, Fernandes et al.
1998).  As a whole, one finds that data in both Table~1 and Figure~1
appear in general agreement with previous evaluations given by Dorman,
Rood \& O'Connel (1993) and Yi et al. (1997), our models being
slightly brighter as a consequence of larger initial He cores (as
discussed in Cassisi et al. 1998).
                                                                       
As expected, Figure~1 shows that the effective temperature of red HB
stars depends on the assumption about the mixing length. This implies
that, at any given temperature, the luminosity of the red portion of
the ZAHB depends on the assumptions on the mixing length, increasing
when the mixing length increases.  This dependence grows with the
stellar metallicity, since at higher metallicities the HB locus moves
toward lower effective temperatures and red HB star are increasingly
affected by external convection, making stellar models more sensitive
to the adopted efficiency of the super-adiabatic convection. As we
will discuss in the next section, this will leave some uncertainty in
theoretical predictions on tilted HB.
               
\section{The CMD morphology of metal-rich HB.}


Making use of model atmospheres by Castelli, Gratton \& Kurucz (1997)
previous results have been translated into CM diagrams. Figure~2 shows
the predicted location of the computed ZAHB sequences in the ($M_{V}$,
B$-$V) and ($M_{V}$, V$-$I) diagrams, for the two choices of the
mixing length. The same figure shows the predicted location of the
corresponding RG branches for $\alpha$=1.6, which gives a reasonable
approximation to the observed RGB colors when adopting the above
atmosphere models (Cassisi et al. 1998).  In the investigated range of
metallicity all the ZAHBs, for each adopted value of the mixing
length, show a canonical tilt, by an amount which is largest at the
highest explored metallicity.  As a matter of fact, the dependence on
the metallicity already discussed in the previous section is now
amplified by the parallel increase of the bolometric correction as the
stellar temperatures decrease.


By relying on a reasonable accuracy of adopted model of atmospheres,
RGB models with a mixing length parameter as low as $\alpha$=1.0 are
ruled out, since they would have exceedingly red colors (see for
example figure 3 in paper I). Thus, if one takes $\alpha$=1.6 for the
actual RGB stars and assumes a common value of the mixing length in
both RGB and HB structures, canonical theory predicts a tilt growing
from $\Delta V$$\sim$0.05 mag when Z=0.002 up to $\Delta V$$\sim$0.2
mag when Z=0.02.

On the other hand, if one assumes a less efficient convection in red
HB stars (i.e. a shorter mixing length) than in RGB structures, this
tilt could increase, as shown in Figure 2.  In such a case one could
invoke either an exceedingly larger cluster age or a not negligible
amount of mass loss to avoid the HB crossing the RGB, an occurrence
which would run against well established observational evidence.
However, such a scenario appears rather artificial, so that we will
base the following discussion on the predictions for the case
$\alpha$=1.6, as a reasonable approximation of actual stellar
structures, suggesting that the above uncertainty on the tilt should
hardly exceed few hundredth of magnitude.

Such a canonical tilt can be obviously observed only in clusters with
a ZAHB covering a suitable range of temperatures, whereas for those
with only red HB one would only detect the sloping red portion of the
branch.  According to the HB models presented here, over the range 0.5
$<$ B$-$V $<$ 1.0, the predicted slope for the solar metallicity is of
the order of $dM_{V}/d(B-V)$$\sim$0.24.  For even redder colors,
models reach their redder HB limit, and luminosity starts increasing
with the mass as early predicted by Caloi, Castellani \& Tornamb\'e
(1978).

One has also to discuss whether evolutionary effects can mask the
discussed ZAHB distribution.  Figure~3 shows the evolutionary paths of
selected HB models for the two given choices on metallicity.  For
Z=0.006 all the evolutionary tracks lie above the ZAHB locus, so that
the lower envelope of the observed distribution keeps being defined by
the ZAHB, thus preserving the sloping behavior discussed in this
section. This is not the case for solar metallicities, where the ZAHB
is crossed by the model having the blue hook just below the point
where its luminosity is at maximum.  However, such an occurrence will
decrease the tilt by less than 0.009 mag, thus again substantially
preserving the predicted canonical tilt.

Let us finally compare our results with evolutionary data recently
presented by Vandenberg et al.  (2000). Inspection of Fig.3 in that
paper (lower panel) discloses that in the case $[\alpha/Fe]=0$
increasing the metallicity causes a tilt to appear, reaching a maximum
value of the order of $\Delta V$$\sim$0.08 mag at the largest
investigated metallicity Z=0.01.  From data shown in our Fig.~2 one
finds $\Delta V$$\sim$0.06 when Z=0.006 and $\Delta V$$\sim$0.18 mag
for solar metallicity. By a linear interpolation on log Z it follows
$\Delta V$$\sim$0.10 at Z=0.01, i.e.  one finds that both results
appear in reasonable agreement, with VandenBerg et al. predicting
marginally smaller tilts.

One can further discuss the origin of the marginal difference, bearing
in mind that the predicted tilt is the combined result of both ZAHB
models and bolometric corrections.  Luckily enough, comparison of
Fig.'s 1 in both papers discloses that we have a quite similar
treatment of external convection, since ZAHB models with the same
chemical composition reach a quite similar minimum temperature. Thus
the difference is not an effect of the already discussed dependence of
the tilt on the mixing length.  On the contrary, there are several
differences both in modeling stellar structures and in color
transformations which could be at the origin of the differences.  We
do not use the same approach for the conductive opacities (Itoh et al.
1983 in our computations against Hubbard \& Lampe 1969), for the
Equation of State (EOS) (Rogers et al. 1996) against an improved
version of VandenBerg (1992) EOS and, in particular our program takes
into account element sedimentation all along the pre-HB evolutionary
phases (Castellani et al. 1997).  Moreover, we use the color -
effective temperature transformations and the BC scale by Castelli,
Gratton \& Kurucz (1997) against Bell \& Gustafsson (1989) and
VandenBerg \& Bell (1985).  When all these differences are taken into
account, one can conclude that the reasonable agreement between
present and VandenBerg et al. (2000) results can be taken as a
encouraging evidence for the reality of tilted HB, as a well
established feature in all modern canonical HB models.

\section{ The case of NGC~6441 and NGC~6388}

In the previous section we have found that canonical HB models predict
moderately tilted HB morphologies. However, in the well known cases of
NGC~6388 and NGC~6441, one finds evidence for a tilt of the order of
$\Delta V$$\sim$0.5 mag, i.e. definitely much larger than predicted on
canonical theoretical grounds. To the light of such an evidence, in
this section let us discuss observational data for these clusters in
order to gain further informations on observational evidences on that
matter.  For the sake of the discussion Figure~4 gives the CM diagram
of NGC~6441 as derived from HST observations (Piotto et al. 2001),
with the clear evidence for the above mentioned tilt.

To be conservative one cannot easily exclude a contribution to the
observed tilt from stars evolved from the blue HB tail, crossing the
CM diagram at (B$-$V)$_{0}$$\sim$0 at luminosities larger than the
ZAHB location.  However, as first noted by Piotto et al. (1997) and
recently reinforced by Pritzl et al. (2001), the cluster appears
peculiar not only for the tilted nature of the whole HB but also for
the slope of the HB red portion. As shown in Fig.~4, the lower
envelope of observed HB stars runs as
$dV/d(B-V)$$\sim$1.5. Comparison with theoretical predictions, as
given in the previous fig.~3, shows that also in this respect
canonical theory can not account for such a feature.

\subsection{Non canonical scenarios}
 
In both NGC~6441 and NGC~6388 the evidence for the tilt is accompanied
by the parallel and unexpected occurrence of extended, extremely blue
HB, against the general red morphology of metal rich HB.  Thus one is
tempted to speculate whether the tilt and the "blue tail" originate
both from a common mechanism.  As a matter of fact, Sweigart \&
Catelan (1998) already suggested three non-canonical scenarios
accounting for these features, as given by i) abnormally high original
Y abundances, ii) a spread in He cores due to stellar rotation, and,
iii) differential deep mixing along the RGB. However, the occurrence
of an extremely high original Y in this cluster has been already
discarded by Layden et al. (1999) on the basis of the R method,
whereas Moehler, Sweigart \& Catelan (1999) found that spectroscopic
results for blue HB stars in these clusters are not consistent with
all these three non-canonical scenarios.

In this context let us drive the attention on an additional
observational evidence running against these three non-canonical
suggestions, as given by the occurrence of a well defined "bump" along
the RG branches of both clusters.  The occurrence of these bumps is
clearly detectable in the original paper by Rich et al. (1997) for
both NGC~6388 and NGC~6441.  Zoccali et al. (1999) have already
discussed the RGB Bump in NGC~6388, showing that the luminosity of the
bump appears in reasonable agreement with theoretical predictions.
Here we suggest that the bump could hardly survive a spread either in
stellar rotation or in deep-mixing, whereas the ``high helium''
scenario runs against the observations.  In fact, data in Figure~4
discloses that, like NGC~6388, also NGC6441 shows a well defined RG
Bump below the HB luminosity level at V$\sim$18.5 mag.  As shown in
the left panel of Figure~5, such a feature appears in fine agreement
with canonical expectations for Y$_{MS}$=0.27.  On the contrary, the
assumption of an original He content as large as Y$_{MS}$=0.43 would
produce quite a small bump, and at too low a luminosity with respect
to the HB luminosity level (Fig.~5, right panel), at odds with the
observations. Thus supporting Layden et al. conclusions on firmer
theoretical basis, i.e. by using the RGB bump which is an
observational features not affected by the typical uncertainties of HB
lifetimes involved in the R-parameter calibration (Cassisi et
al. 1998).


\subsection{Spread in metallicity}

In the recent literature it has repeatedly suggested that the peculiar
HB in the two clusters could be connected with a spread in the
metallicity of cluster stars (Piotto et al. 1997, Sweigart 2000,
Pritzl et al. 2001).  To discuss this point we can take advantage of
the well developed blue HB of NGC~6388, covering the abrupt decrease
of magnitude defined as the HB-turning down (HB-TD) in Brocato et
al. (1998). According to the discussion given in that paper one can
safely assume the HB-TD as indicator of intrinsic CM location,
independently of HB metallicities. This allow a meaningful comparison
of the CM diagram of NGC~6388 with similar diagram but for less metal
rich clusters. To perform such a comparison we choose from the Padova
HST snapshot data-base the CM diagram for two clusters showing well
developed HB blue tail, namely NGC 7099 ([Fe/H]=$-$2.12) and NGC 2808
([Fe/H]=$-$1.15), both metallicities in Zinn and West scale.

The results of such a procedure, as disclosed in Figure~6, allow to
put firm constraints on the presence of metal poor stellar populations
in NGC~6388. If the blue HB stars in this cluster are originated from
metal poor progenitors, one should expect along the luminous portion
of the RGB as many RGB stars as in the blue HB phases, since the
predicted ratio R of HB to RG stars above the HB luminosity level is
of the order of unity (see e.g. Zoccali et al. 2000).  Since the blue
HB of NGC~6388 contains 202 stars, one can be confident that the lack
of observational evidence of RGB stars in the expected regions in the
NGC~6388 diagram excludes the occurrence of metal poor giants with the
given metallicities. One concludes that in NGC~6388 blue HB stars must
be originated from RG progenitors with metallicity sensitively larger
than the metallicity of NGC 2808, namely larger than [Fe/H]=$-$1,
leaving little room to the hypothesis of a spread in metallicity.



\subsection{The role of differential reddening}

In order to save canonical predictions it has already suggested that
differential reddening in these heavy reddened clusters could play a
relevant role.  The occurrence of differential reddening in NGC~6441
has been already suggested by Piotto et al. (1997).  Layden et
al. (1999) have recently discussed their ground-based CCD photometry
of NGC~6441, confirming that cluster stars are affected by a rather
strong differential reddening. By noticing that the slope of the HB
runs according to the reddening vector, these authors conclude that
differential reddening should partially contribute to the observed
slope.  To look into this problem we took advantage of HST
observation, looking for the features of CM diagrams in selected
cluster areas. As reported in Figure~7, we extracted from the
36.4$\times$36.4 arcsec$^{2}$ image of the planetary camera 16
(9.1$\times$9.1 arcsec$^{2}$) subimages, comparing the CM diagrams of the
various areas with the RGB ridge line of area \#6 taken as a
reference. The size of each small area is chosen in order to have a
reasonable number of stars in the corresponding RGB.


Inspection of data in Fig.~7 reveals indeed some relevant features. As
a first point, one finds clear evidence for the occurrence of rather
strong reddening differences, as disclosed, e.g., by the extreme case
of area 4, where the CM diagram shows a color shift as large as about
$\Delta$(B$-$V)$\approx$0.15 mag with respect to the reference
diagram.  This result is in good agreement with the maximum
differential reddening estimated by Heitsch \& Richtler (1999) from
ground-based observations of the same cluster.

Moreover, the reference diagram shows a narrow RG sequence, suggesting
that the observed width of the cluster RG branch could be just an
artifact of differential reddening across the cluster. If this is the
case, this feature, together with the presence of the blue tail,
provides a further constraint against the hypothesis of a spread of
metallicity in the cluster stars.

A differential reddening of $\Delta E(B-V)$$\sim$0.1 appears large
enough to produce the observed red-ward slope of the HB.  To
substantiate such a scenario, we performed a numerical experiment,
applying a random differential reddening by $\Delta{E(B-V)}_{max}=0.1$
to the observed CM diagram of 47~Tuc, in such a way to reproduce the
color dispersion observed in the RGB of NGC~6441.  The result of this
experiment, as presented in Figure~8, shows the impressive similarity
between NGC~6441 (Fig.~4) and the "reddened" 47~Tuc.  As a relevant
point, neither in NGC 6441 nor in the "reddened" 47~Tuc the HB slope
is exactly along the adopted reddening vector. This occurance is the
natural consequence of randomly reddening a HB which is less sloped
than the reddening vector. To get the maximum change in slope one
shoud apply the minimum amount of reddening to the hotter red HB stars
and a larger amount to the cooler red HB stars.  Thus, no
matter how one builds up the HB in this way the resulting slope of the
lower envelope will be less than that of the reddening vector.

While differential reddening {\em could} convert a 47~Tuc-like HB into one like
the red HB of NGC~6441, it turns out that is not the case.  To see this
one can note the thin RGB ($\Delta (B-V)$$<$0.1 mag) shown by the reference
diagram \#6 in Fig.~7 (but see also diagrams 5, 10, 13, 14).  This
shows that there is not substantial differential reddening within those
subsamples. Yet one sees that the red HB in the same diagram is
clearly sloping down with increasing B$-$V color.  
Indeed, the HB in each subsample has a slope consistent with that of
the global sample. Since the slope of each subsample is already fairly
large, differential reddening will stretch the HB without
substantially changing the slope.  Since the large slope is present in
even subsample with small differential reddening, it must result from
a real behavior of cluster HB stars.

Finally, let us note that by applying the same analysis to WFPC2-HST
observation of NGC~6388, we derive results on the differential
reddening which are consistent with the previous discussion on
NGC~6441.

\section{Final remarks}  

Heitsch \& Richtler (1999) have presented new photometry for five
globular clusters in the inner Galaxy concluding that differential
reddening is indeed responsible for the sloped red HB morphologies at
least in the case of NGC 5927. However, we have already advanced the
suggestion of a real slope at least in NGC~6441. An inspection of the
CM diagrams for the metal rich globulars in the Padova HST snapshot
data-base (Piotto et al. 2001) plotted in Fig.~9 reveals that all the
clusters more metal rich than 47Tuc, but NGC 6624, show red HB stars
with a slope close to $dV/d(B-V)$=1.5 indicated by the continuous
line.  As for NGC 6624 a best agreement is reached with the dotted
line, for a slope of the order of $dV/d(B-V)$=1.0. Note that a
similar slope can be clearly detected in the beautiful V, V$-$I
color-magnitude diagram recently presented by Heasley et al. (2000).
The HB of 47Tuc has a shallow slope, close to the $dV/d(B-V)$=0.2
expected from the theoretical models.

Since all the clusters with sloping HB have also a substantial amount
of reddening it appears difficult to ascertain the role of possible
differential reddening. In this context, the reduced slope in NGC 6624
could originated either from a smaller differential reddening or from
a smaller metallicity ([Fe/H]=$-$0.63$\pm$0.09) as recently discussed
by Heasley et al. (2000).  However, the evidence for sloping red HB in
clusters like NGC 6356 or NGC 6624, with a rather well defined RG
sequence, suggest that the sloping red HB is a real feature of metal
rich clusters. If this is the case, one should conclude that sloping
HBs {\it are not necessarily correlated with the occurrence of blue
tails}.

Comparison with theoretical predictions in the previous Fig.~2 shows
that such a slope can be attained by theoretical models only for the
shorter mixing length value we already regarded as largely
improbable. If this slope is due to intrinsic properties of HB stars,
one should conclude that the observations tell us that when the
increasing metallicity pushes the HB at larger B$-$V colors, then the
red HB slopes down, for unknown reasons.  One would naturally account
for such an occurrence by increasing the dependence of the bolometric
correction on the temperature, but by an amount that is believed to be
beyond current uncertainties. Therefore this remains an open question
requiring further investigations.


In this context one should finally notice that current theoretical
procedures adopted in producing HB models start with equilibrium ZAHB
models which neglect all the phases of pre-HB evolution as well as the
phases of readjustment of CNO elements across the H burning shell
(see, e.g. Castellani, Chieffi \& Pulone 1989).  The point is to
understand if a pre-ZAHB evolution can play a role in the observed HB
morphologies. However, a close inspection on the recent models
performed to follow the evolution through the He-flash by Brown et
al. (2001) and preliminary computations by Piersanti L. (private
communication) do not support such a possibility.

\section{Acknowledgements}

We are grateful to D. VandenBerg for providing his theoretical ZAHB
models in ascii format. It is a pleasure to thank the Referee for
his invaluable work in providing criticism but also very
constructive comments and suggestions which greatly improved 
the final version of the paper. This work is supported by the
Italian Ministry of University, Scientific Research and Technology
(MURST) within the Cofin2000-Project: {\sl Stellar Observables of
cosmological relevance} and the Cofin1999-Project: {\sl Stellar
Dynamics and Stellar Evolution in Globular Clusters}. Partial support
by the Agenzia Spaziale Italiana is also acknowledged.


\clearpage

\clearpage
 
\figcaption[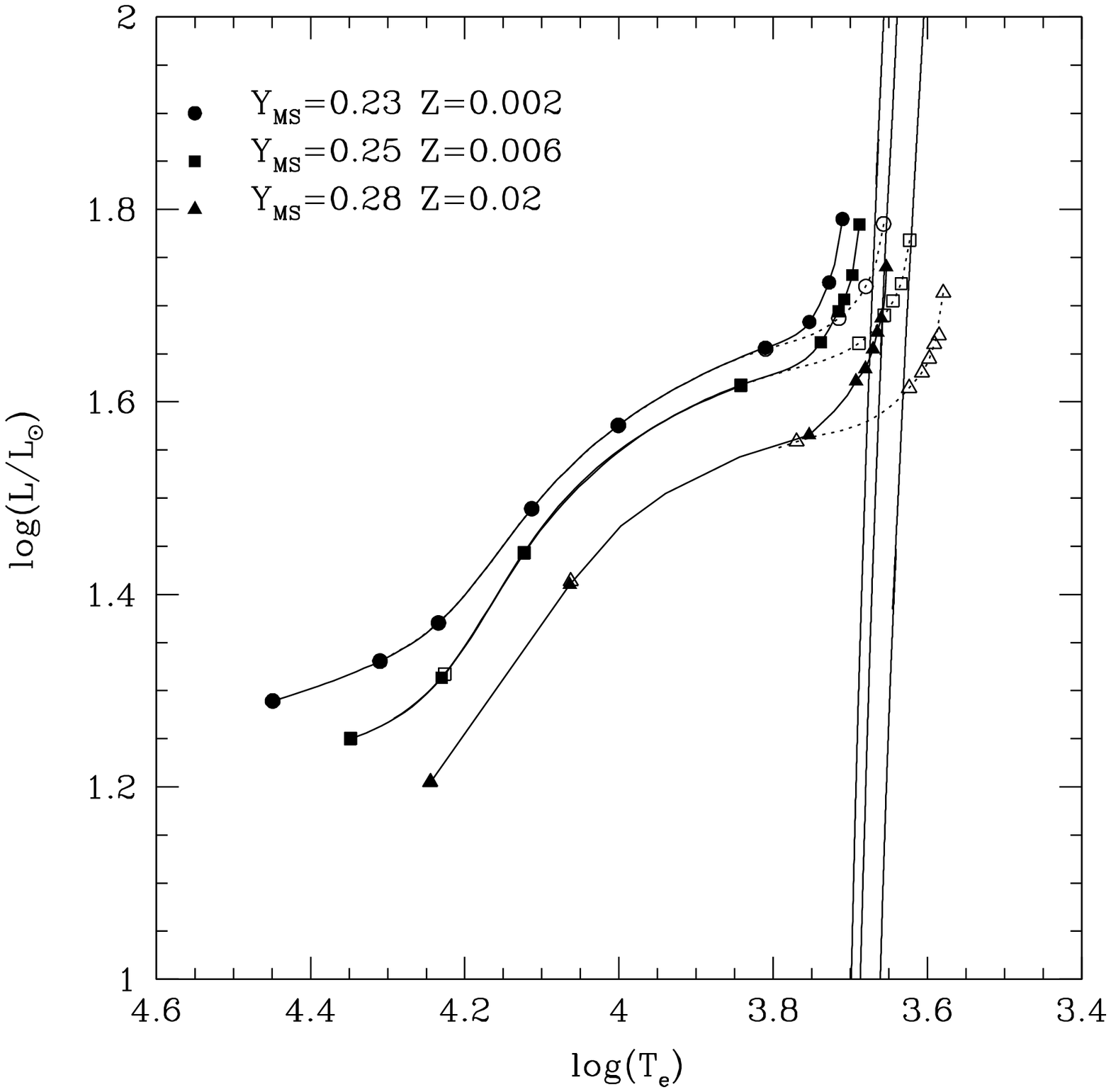]
{ZAHB loci in the theoretical log(L/L$_{\odot}$),
log(T$_{e}$) plane for the labelled chemical compositions (Y$_{MS}$,
Z) and for two different assumptions about the mixing length parameter
($\alpha$=1.0, dotted lines and open symbols, and 1.6 solid lines and
filled symbols).  Symbols along the ZAHB give the location of models
with masses 0.505 (0.508 for Z=0.002), 0.53, 0.55, 0.58, 0.60, 0.63,
0.65, 0.70 and 0.90M$_{\odot}$ respectively. The location of the
corresponding Red Giant Branches in the case $\alpha$=1.6 is also
shown, with metallicity increasing left to the right.
\label{fig1}}

\figcaption[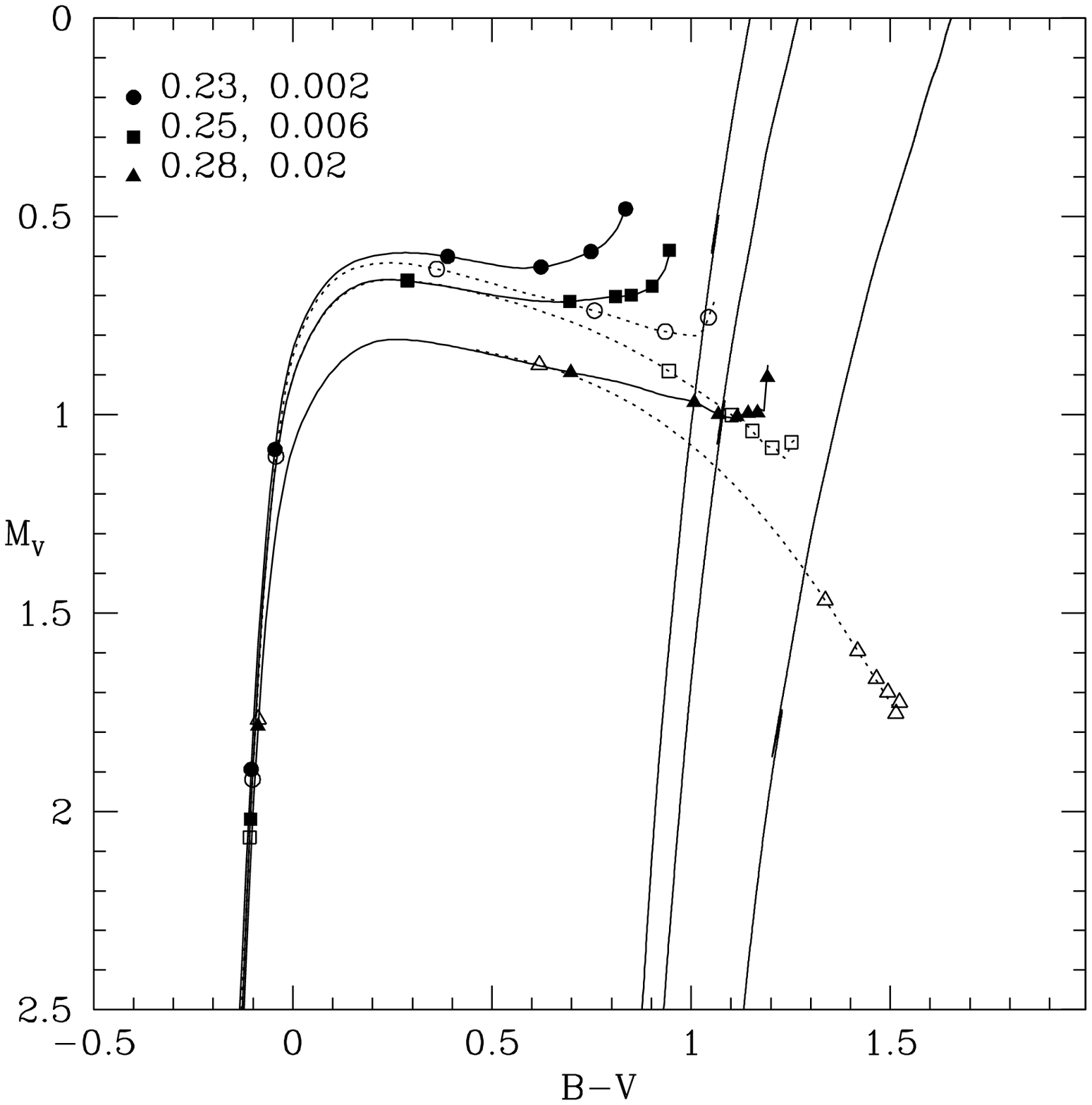,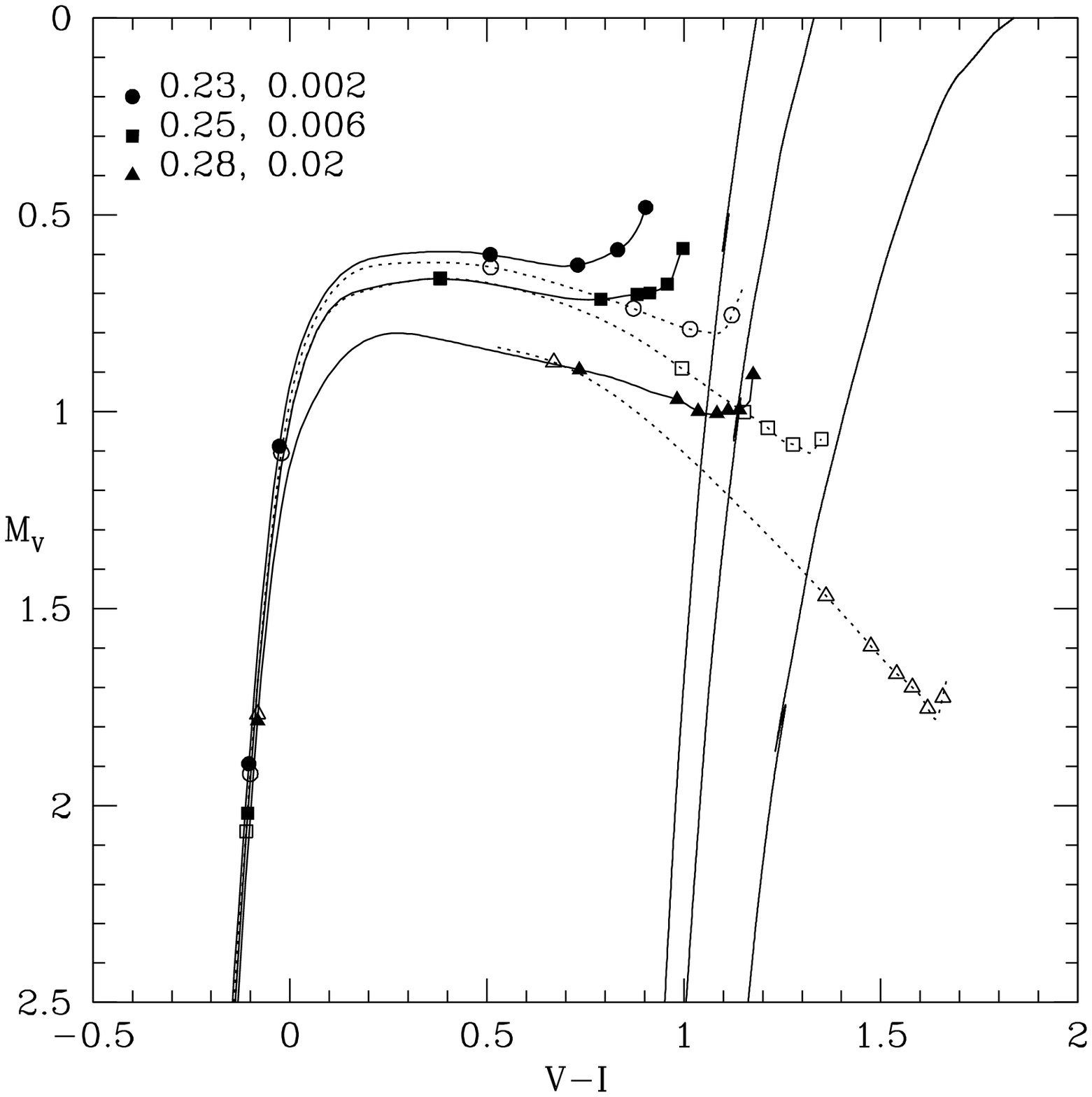]
{ZAHBs and RG branches as in Figure~1 but in the observational 
(M$_V$, B$-$V) and (M$_V$, V$-$I) planes. \label{fig2}}

\figcaption[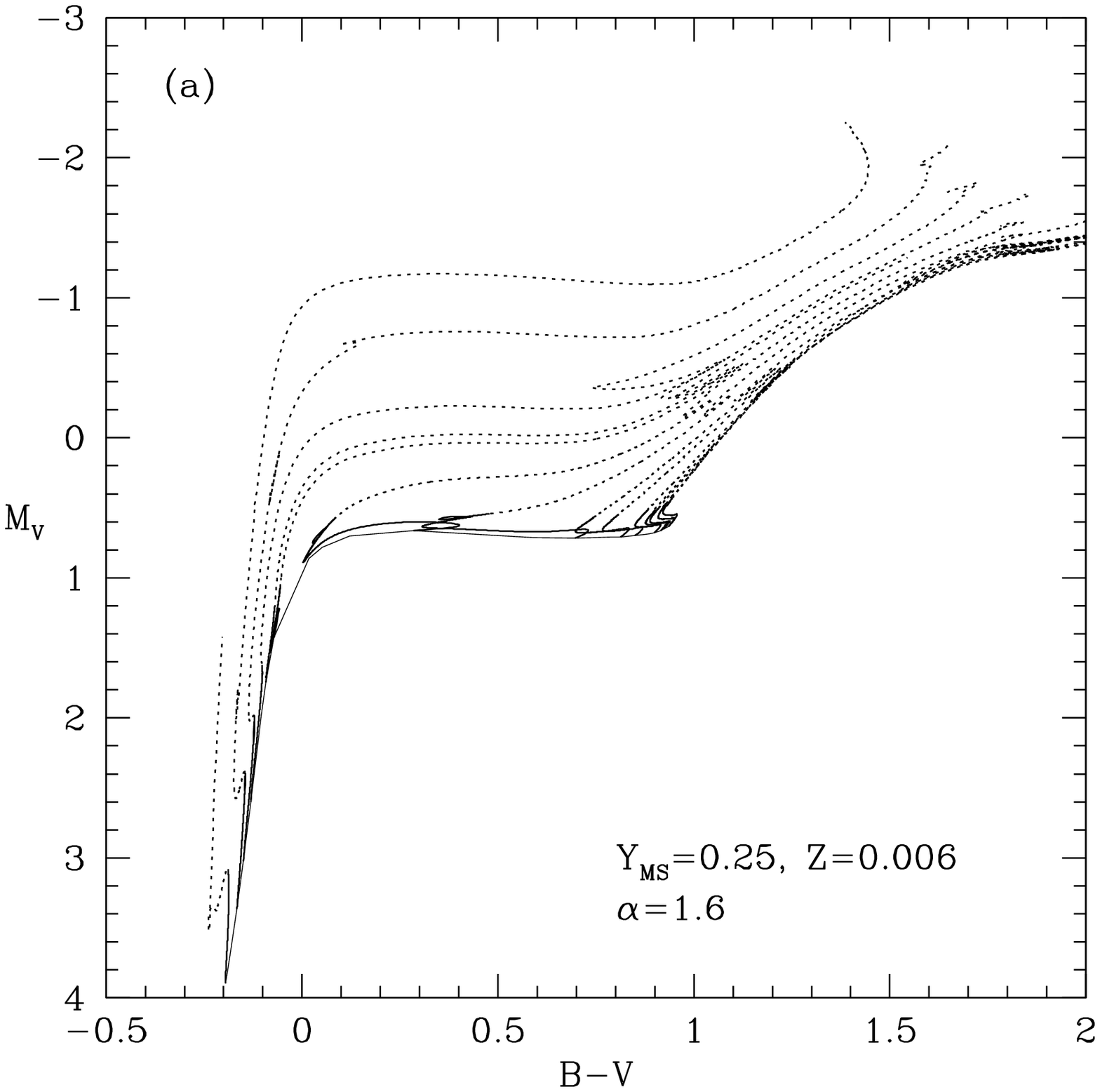,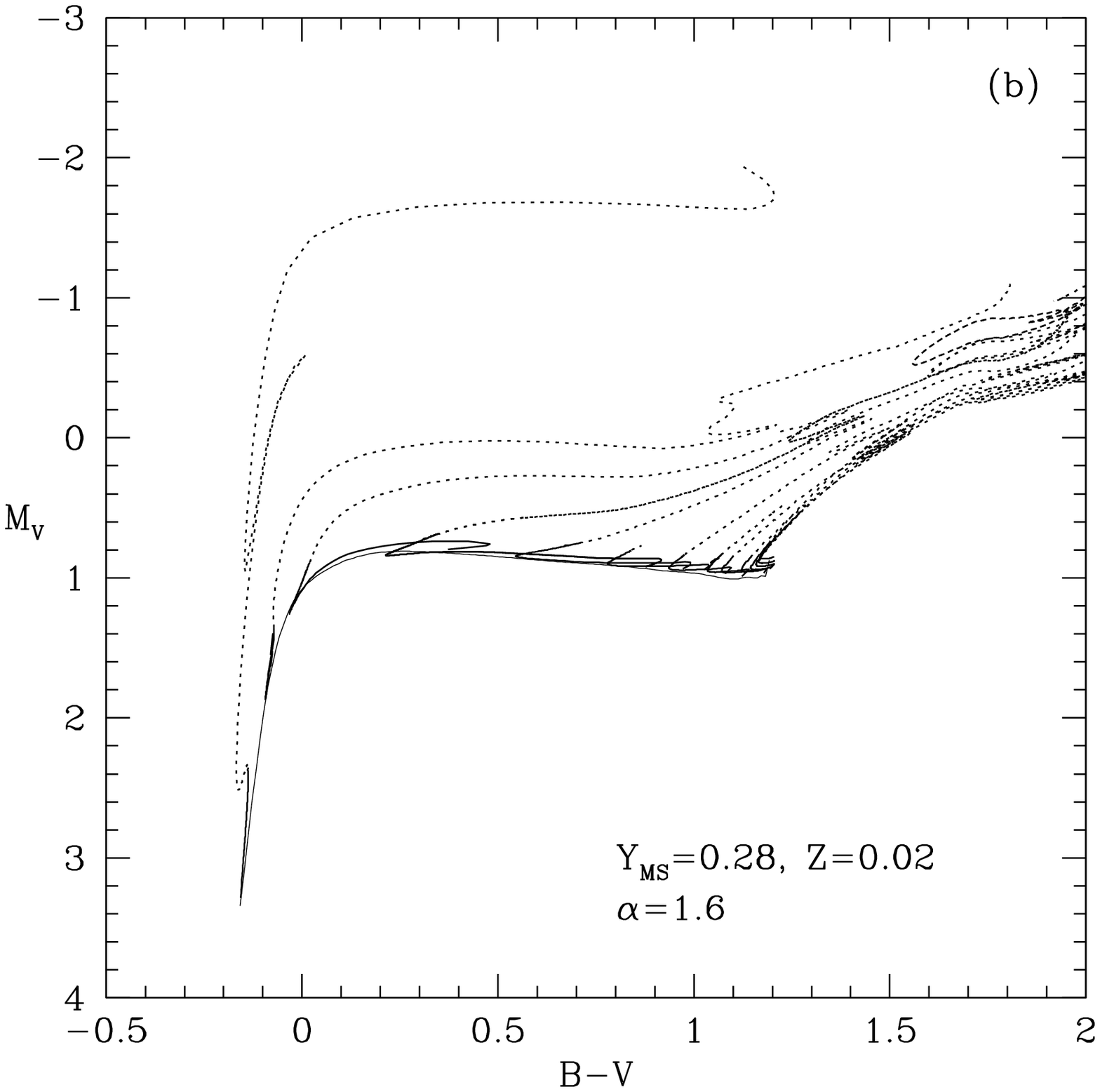]
{{\sl Panel (a)} A selected sample of evolutionary tracks 
with Z=0.006, Y$_{MS}$=0.25 and $\alpha$=1.6 and 
the corresponding ZAHB in the observational (M$_{V}$, B$-$V) plane. 
Dotted lines refer to the final phases of He-burning, when 
the central He abundance is lower then Y$_{c}$=0.1. 
{\sl Panel (b)} As in {\sl Panel (a)} for Z=0.02 and
Y$_{MS}$=0.28. \label{fig3}}

\figcaption[raimondo.f4.ps]
{The (V,B$-$V) diagram of NGC~6441. The arrow shows 
the observed location of the RGB bump.  \label{fig4}}

\figcaption[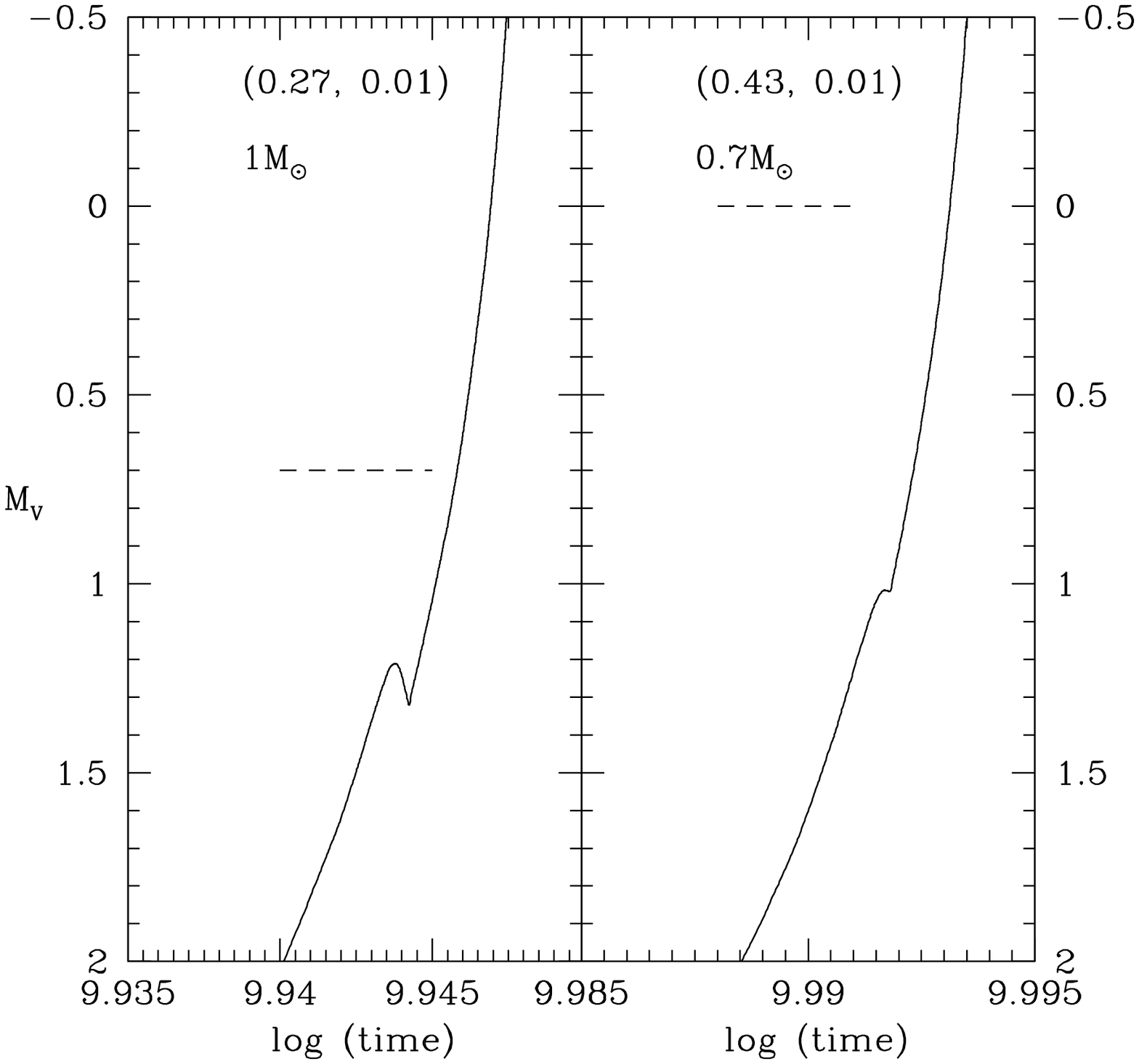]
{Theoretical RG branches for a cluster age of the order of
10 Gyr, Z= 0.01 and two different assumptions
on the original helium content Y$_{MS}$= 0.27 (M= 1.0M$_{\odot}$) and 
Y$_{MS}$= 0.43 (M=0.7M$_{\odot}$). 
The dashed lines show the corresponding predictions 
about the HB luminosity level. \label{fig5}}

\figcaption[raimondo.f6a.eps,raimondo.f6b.eps]
{{\sl Left Panel:} The CM digram of NGC 7099 is forced 
to match the blue side of the HB of NGC~6388. 
{\sl Right Panel:} The same for the more metal rich cluster NGC 2808. 
\label{fig6}}

\figcaption[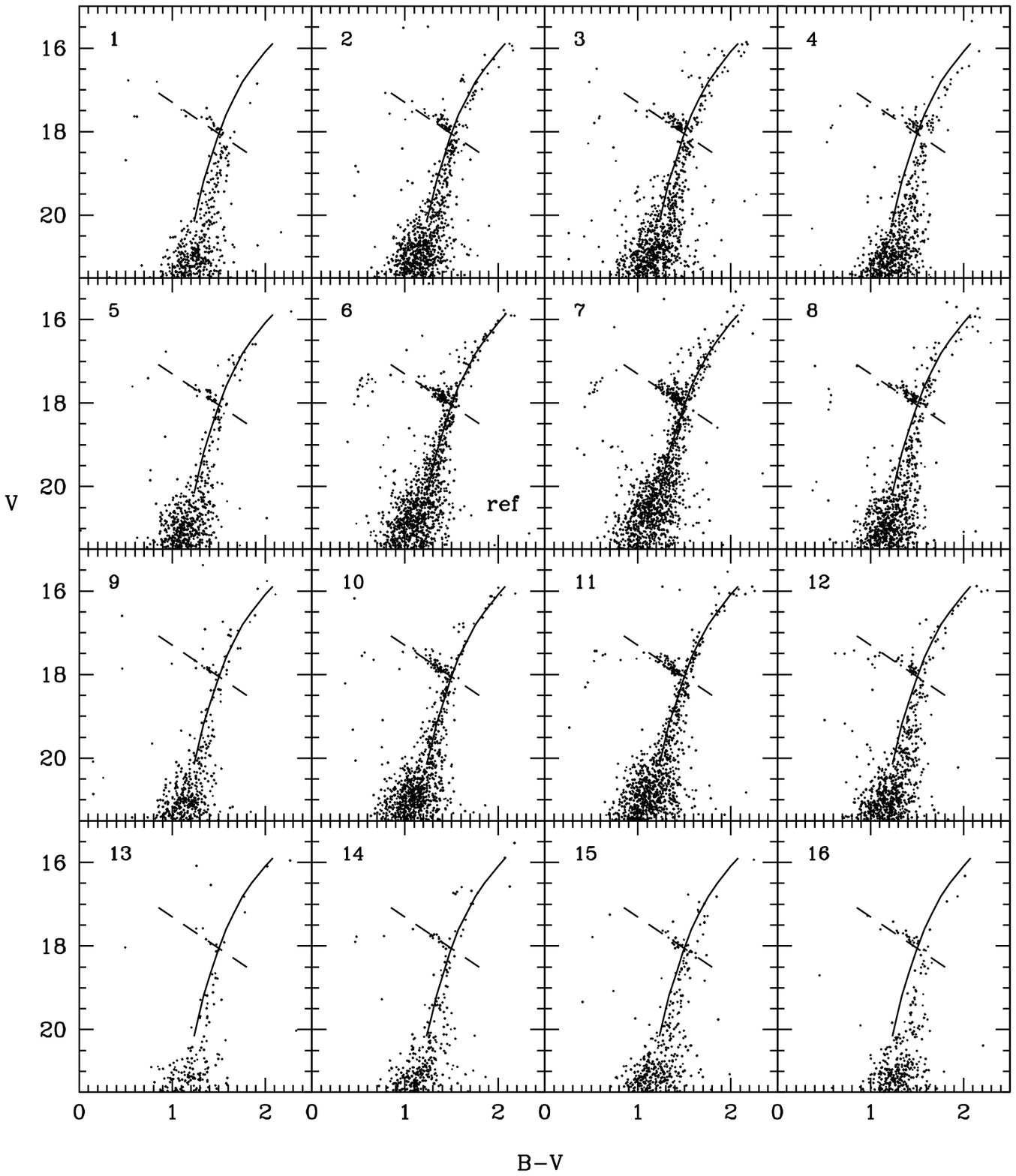]
{CM diagrams in 16 areas (9.1$\times$9.1 arcsec$^{2}$) of the planetary
camera (PC) subimage of NGC~6441. The RGB ridge line (solid line) 
of the reference diagram in box \#6 is over-plotted on the other diagrams. 
The dashed line in each subimage is sloped as $dV/d(B-V)$=1.5.
\label{fig7}}

\figcaption[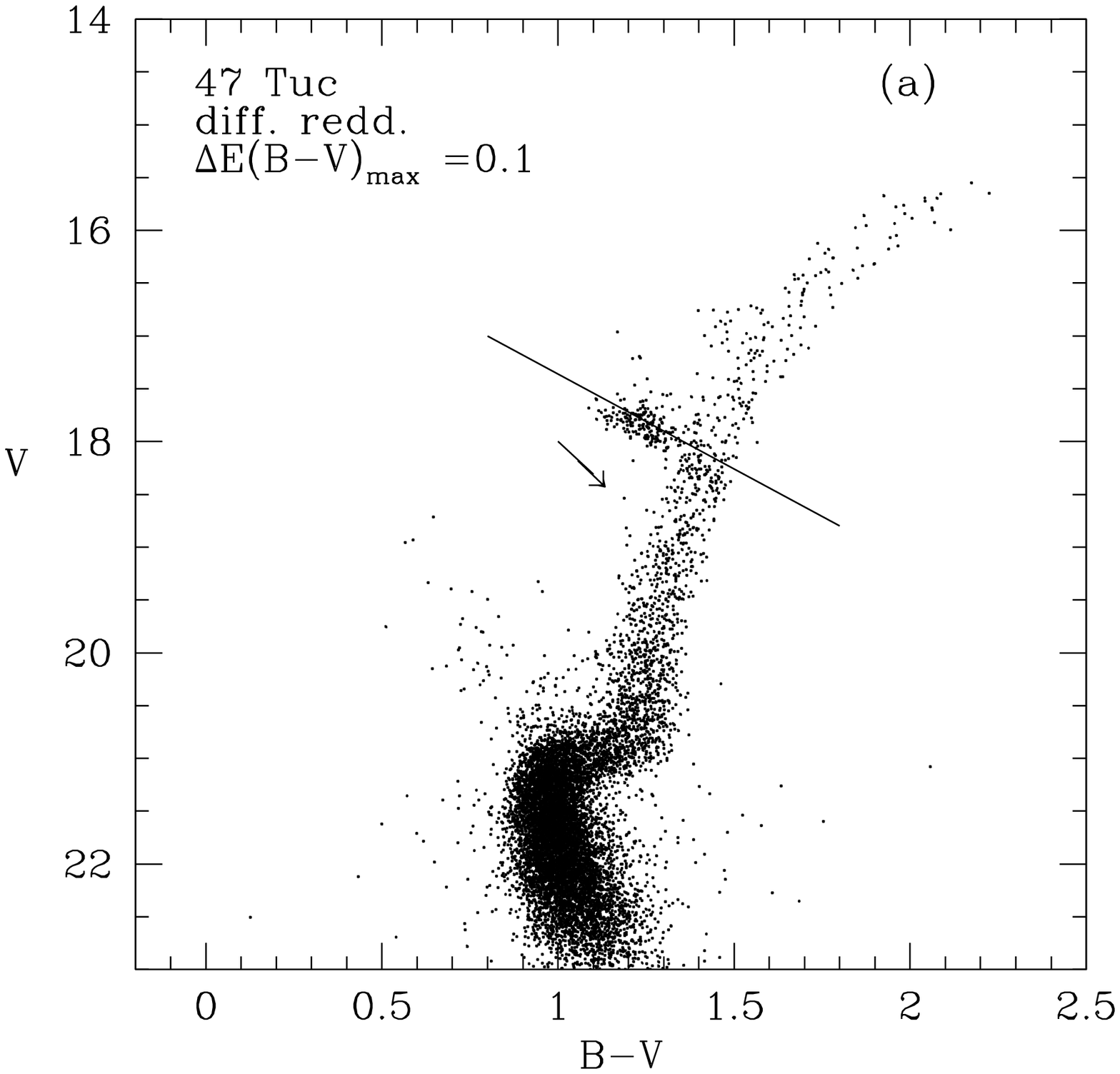,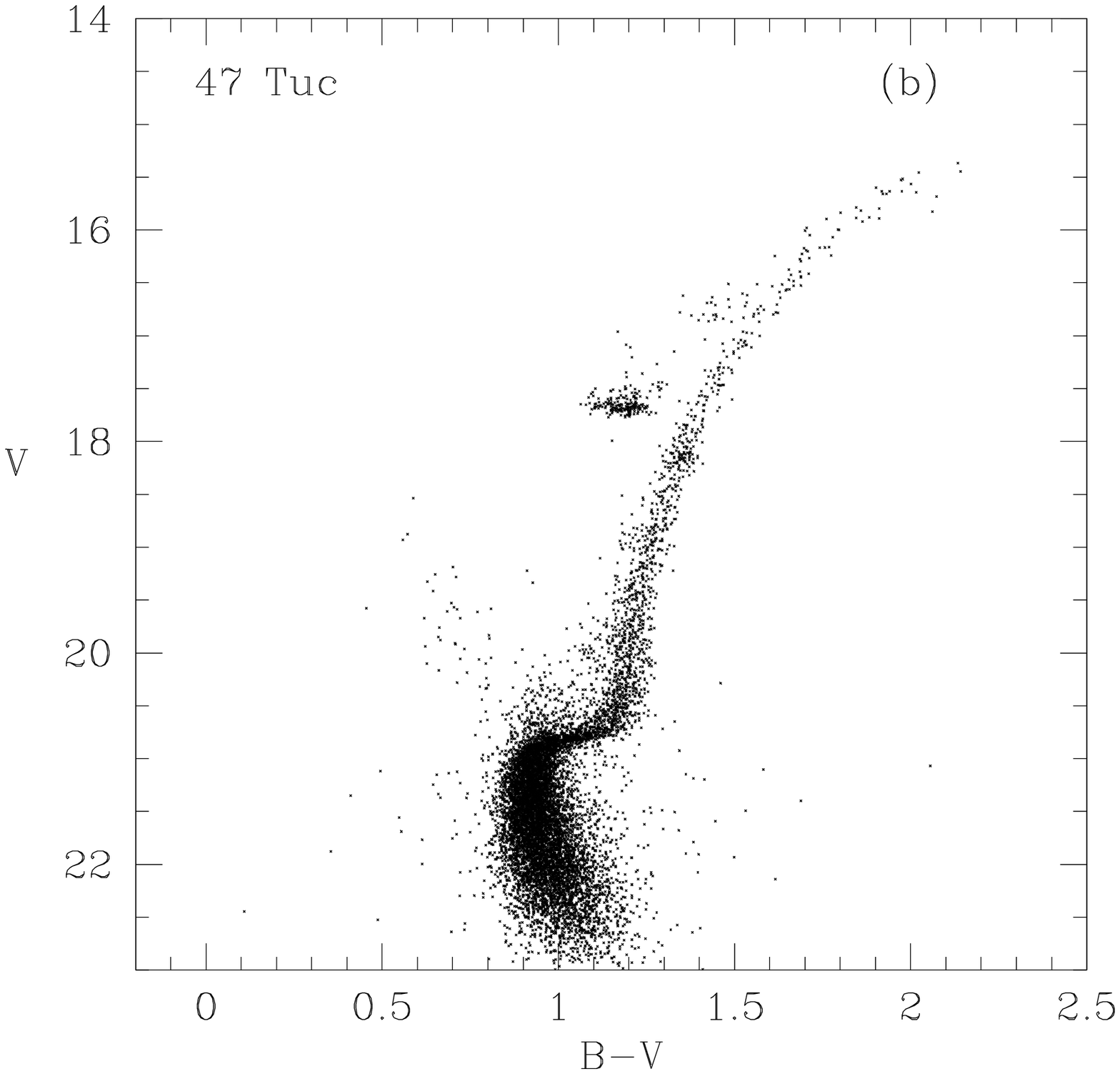]
{{\sl Panel (a)} The CM diagram of 47~Tuc arbitrarily shifted by 3.6
mag in V and 0.38 mag in B$-$V to match the NGC~6441 diagram, with an
artificial differential reddening (see text for more details).  The
plotted arrow represents the slope of the reddening vector (Cardelli,
Clayton \& Mathis 1989) and the solid line is drawn by assuming a
slope of $dV/d(B-V)$=1.5. {\sl Panel (b)}: The original CM diagram
of 47~Tuc. \label{fig8}}

\figcaption[raimondo.f9.ps]
{HBs of six well populated clusters in Piotto et al. (2001)
data-base.  The solid line is drawn by assuming a slope of
$dV/d(B-V)$=1.5 and the dotted one by $dV/d(B-V)$=1.0 (see text).
 The arrow in the upper left panel is the
reddening vector ($R_{V}=3.1$). Metallicity and reddening are 
from Harris (1999). \label{fig9}}

\clearpage

\begin{deluxetable}{ccccccc}
\small
\tablenum{1}
\tablecaption{Selected evolutionary quantities at the He ignition
for different assumptions on the RGB progenitor and on the chemical composition.}
\tablewidth{450pt}
\tablehead{
\colhead{$Y_{MS}$} & \colhead {Z} & \colhead {$M_{RGB}(M_\odot)$} & \colhead {$M_{cHe}(M_\odot)$} &
\colhead {$Y_{sup}$} & \colhead {$Age^{tip} (Gyr)$} & \colhead {$\log(L/L_{\odot}^{tip})$}}
\startdata
0.23 & 0.002 & 0.9  & 0.501  & 0.23   &  10.23      & 3.42    \\
0.25 & 0.006 & 0.9  & 0.492  & 0.26   &  11.75      & 3.44    \\
0.28 & 0.02  & 1.0  & 0.481  & 0.29   &  11.22      & 3.47    \\
\enddata
\end{deluxetable}                                                                                   

\clearpage


\clearpage

\begin{thebibliography}{}
\bibitem[]{bel89} Bell, R. A., Gustafsson, B. 1989, MNRAS 236, 653
\bibitem[Bono et al.(1997)]{bono97} Bono, G., Caputo, F.,  Cassisi, S., Castellani, V., 
Marconi, M. 1997, ApJ, 489, 822
\bibitem[Borissova et al.(1999)]{bor99} Borissova, J., Catelan, M., Ferraro, F. R., Spassova, N. 
Buonanno, R., Iannicola, G., Richtler, T., Sweigart, A. V. 1999, A\&A, 343, 813
\bibitem[Brocato et al.(1998)]{bro98} Brocato, E., Castellani, V., Scotti, G. A., Saviane, I., Piotto, G., Ferraro, F. R., 1998, A\&A 335, 929
\bibitem[Brocato et al.(1999)]{bro99} Brocato, E., Castellani, V., Raimondo, G., Walker, A. R. 
1999, ApJ, 527, 230 (Paper I)
\bibitem[Brown et al.(2001)]{brow01} Brown T. M., Sweigart A. V., Lanz T., Landsman W. B., Hubeny I., 2001, astro-ph 0108040
\bibitem[Caloi, Castellani and Tornamb\`e]{cal78} Caloi, V., Castellani, V., Tornamb\`e, A., 1978, A\&AS 33, 169
\bibitem[Cardelli, Clayton and Mathis(1989)]{car89} Cardelli, J. A., Clayton, G. C., \& Mathis, J. S. 1989, 
ApJ, 345, 245
\bibitem[Cassisi et al.(1998)]{san98} Cassisi, S., Castellani, V., Degl'Innocenti, S., Weiss, A. 
1998, A\&AS, 129, 267
\bibitem[CAssisis et al.(1999)]{san99} Cassisi, S., Castellani, V., Degl'Innocenti, S., Salaris, 
M., Weiss, A. 1999, A\&AS, 134, 103
\bibitem[Castellani, Chieffi and Pulone(1989)]{cast89} Castellani V., Chieffi A., Pulone L., 1989, 344, 239
\bibitem[Castellani et al.(1997)]{cast97} Castellani V., Ciacio F., Degl'Innocenti S., Fiorentini G. 1997, A\&A 322, 801
\bibitem[Castelli, Gratton and Kurucz(1997)]{cas97} Castelli, F., Gratton, R.G., Kurucz, R. L. 1997, A\&A, 318, 
841
\bibitem[Chaboyer et al.(1998)]{cha98} Chaboyer, B., Demarque, P., Kernan, P. J., Krauss, L. M. 
1998, ApJ, 494, 96
\bibitem[Dorman, Rood and O'Connel(1993)]{dor93} Dorman, B., Rood, R. T., \& O'Connel, R. W. 1993, ApJ, 419, 
596
\bibitem[Fagotto et al.(1994)]{fag94} Fagotto, F., Bressan, A., Bertelli, G., Chiosi, C. 1994, 
A\&AS, 105, 29
\bibitem[Fernades et al.(1998)]{fer98} Fernandes, J., Lebreton, Y., Baglin, A., Morel, P. 1998, 
A\&A, 338, 455
Cassisi, S., Petro, L. D., Saha, A., Shara, M. M. 1998, ApJ, 507, 818
\bibitem[]{gir00} Girardi, L., Bressan, A., Bertelli, G., Chiosi, C. 2000, 
A\&AS, 141, 371
\bibitem[]{grat97} Gratton, R. G., Fusi Pecci, F., Carretta, E., Clementini, 
G., Corsi, C. E., Lattanzi, M. G. 1997, ApJ, 491, 749
\bibitem[]{ith83} Itoh, N., Mitake, S., Iyetomi, H., Ichimaru, S. 1983, ApJ 273, 774
\bibitem[]{har99} Harris W. E., 1999, http://physun.physics.mcmaster.ca/Globular.html
\bibitem[]{hea00} Heasley, J. N., Janes, K. A., Zinn, R., Demarque, P., Da Costa, G. S., Christian, C. A., 2000, AJ, 120, 879
\bibitem[]{hei99} Heitsch, F., \& Richtler, T. 1999, A\&A, 347, 455
\bibitem[]{hor92} Horch, E., Demarque, P., \& Pinsonneault, M. 1992, ApJ, 
388, 53
\bibitem[]{hub69} Hubbard, W. B., Lampe, M. 1969, ApJS 18, 297
\bibitem[]{lay99} Layden, A. C., Ritter, L. A., Welch, D. L., Webb, T. M. A. 
1999, AJ, 117, 1313
\bibitem[]{moeh99} Moehler, S., Sweigart, A. V., \& Catelan, M. 1999, A\&A, 
351, 519
\bibitem[]{orto00} Ortolani, S., Momany, Y., Barbuy, B., Bica, E., Catelan, M., 2000, A\&A, 362, 953
\bibitem[]{pei95} Peimbert, M. 1995, in The Light Element Abundances, ed. P. 
Crane (Berlin: Springer-Verlag), p165
\bibitem[]{pio97} Piotto, G., Sosin, C., King, I. R., Djorgovski, S. G., 
Rich, R. M., Dorman, B., Renzini, A., Phinney, S., Liebertet, J. 1997, 
in Advances in Stellar Evolution, ed. R.T. Rood \& A. Renzini (Cambridge 
Univ. Press), p84
\bibitem[]{pio01} Piotto, G., et al. 2001, in preparation
\bibitem[]{prit01} Pritzl, B., Smith, H. A., Catelan, M., Sweigart, A. V., 2001, AJ, 122, 2600
\bibitem[]{ric97} Rich, R.M., Sosin, C., Djorgovski, S. G., Piotto, G., King, 
I. R., Renzini, A., Phinney, E. S., Dorman, B., Liebert, J., Meylan, G. 
1997, ApJ, 484, L25
\bibitem[]{rog96} Rogers F.J., Swenson F.J., Iglesias C.A. 1996, ApJ 456, 902
\bibitem[]{sal97} Salaris, M., \& Weiss, A. 1997, A\&A, 327, 107
\bibitem[]{swe00} Sweigart, A.V., 2000, in ``Mixing and Diffusion in Stars: Theoretical Predictions and Observational Constraints'', 24th meeting of the IAU, Joint Discussion 5, August 2000, Manchester, England
\bibitem[]{swe98} Sweigart, A.V., \& Catelan, M. 1998, ApJ, 501, L63
\bibitem[]{van92} VandenBerg, D. A., 1992, ApJ 391, 685 
\bibitem[]{van85} VandenBerg, D. A., Bell, R. A. 1985, ApJS 58, 561
\bibitem[]{van00} VandenBerg, D. A., Swenson, F. J., Rogers, F. J., Iglesias, C. A.,
 Alexander, D. R. 2000, ApJ 532, 430
\bibitem[]{wal98} Walker, A. R. 1998, AJ, 116, 220
\bibitem[]{yi97} Yi, S., Demarque, P., \& Kim, Y.-C. 1997, ApJ, 482, 677
\bibitem[]{zoc99} Zoccali, M., Cassisi S., Piotto, G., Bono G., Salaris, M.
1999, ApJ 518, L49
\bibitem[]{zoc00} Zoccali, M., Cassisi S., Bono G., Piotto, G., Rich, R., M., Djorgovski, S. G.,
2000, ApJ 538, 289
\end{thebibliography}
\end{document}